# Highly anisotropic Drude-weight-reduction and enhanced linear-dichroism in van der Waals Weyl semimetal $T_d$-MoTe$_2$ with coherent interlayer electronic transport

*Bo Su, Weikang Wu, Jianzhou Zhao, Xiutong Deng, Wenhui Li, Shengyuan A. Yang, Youguo Shi, Qiang Li, Jianlin Luo, Genda Gu, and Zhi-Guo Chen\**

Dr. B. Su, X. Deng, Dr. W. Li, Prof. Y. Shi, Prof. J. Luo, Prof. Z.-G. Chen
Beijing National Laboratory for Condensed Matter Physics
Institute of Physics, Chinese Academy of Sciences
Beijing 100190, China
Email: zgchen@iphy.ac.cn

Prof. W. Wu
Key Laboratory for Liquid-Solid Structural Evolution and Processing of Materials,
Ministry of Education, Shandong University,
Jinan 250061, China

Prof. J. Zhao
Co-Innovation Center for New Energetic Materials
Southwest University of Science and Technology
Mianyang, Sichuan 621010, China

Prof. S. A. Yang
Research Laboratory for Quantum Materials
Singapore University of Technology and Design
Singapore 487372, Singapore

Prof. Q. Li, Prof. G. Gu
Condensed Matter Physics and Materials Science Department,
Brookhaven National Lab,
Upton, New York 11973, USA

Prof. Q. Li
Department of Physics and Astronomy,

Stony Brook University,

Stony Brook, New York, 11794 USA

Prof. Y. Shi, Prof. J. Luo, Prof. Z.-G. Chen

Songshan Lake Materials Laboratory

Dongguan, Guangdong, 523808, China

Keywords: type-II Weyl semimetal, anisotropic charge dynamics, optical-absorption anisotropy, optical spectroscopy

**Weyl semimetal (WSM) states can be achieved by breaking spatial-inversion symmetry or time reversal symmetry. However, the anisotropy of the energy reduction contributing to the emergence of WSM states has seldom been investigated by experiments. A van der Waals metal $MoTe_2$ exhibits a type-II WSM phase below the monoclinic-to-orthorhombic-phase-transition temperature $T_c \sim 250$ K. Here, we report a combined linearly-polarized optical-spectroscopy and electrical-transport study of $MoTe_2$ at different temperatures. The Drude components in the $a$-axis, $b$-axis and $c$-axis optical conductivity spectra, together with the metallic out-of-plane and in-plane electrical resistivities, indicate the coherent inter-layer and in-plane charge transports. Moreover, the Drude weight in $\sigma_{1a}(\omega)$, rather than the Drude weights in $\sigma_{1b}(\omega)$ and $\sigma_{1c}(\omega)$, decreases dramatically below $T_c$, which exhibits a highly anisotropic decrease in its Drude weight and thus suggests a strongly anisotropic reduction of the electronic kinetic energy in the WSM phase. Furthermore, below $T_c$, due to the in-plane anisotropic spectral-weight transfer from Drude component to high-energy region, the in-plane inter-band-absorption anisotropy increases remarkably around 770 meV, and has the largest value (~ 0.68) of normalized linear dichroism among the reported type-II WSMs. Our work sheds light on seeking new WSMs and developing novel photonic devices based on WSMs.**

# 1. Introduction

Weyl semimetals (WSMs), which host pairs of Weyl cones in the bulk and topologically protected Fermi arcs on the surface,[1-19] have generated enormous interest because they not only can offer a rich avenue for exploring novel quantum phenomena including quantum anomalous effect and chiral magnetic effect,[20-28] but also can provide a fertile ground for developing a new generation of high-performance photonic devices.[29-34] Therefore, searching for new WSMs is one of the cutting-edge research areas in condensed matter science. It is well known that breaking spatial-inversion symmetry or time reversal symmetry can result in WSM states.[1-19] However, the anisotropy of the driving force for the occurrence of breaking the spatial-inversion symmetry or time reversal symmetry in WSMs, which would offer a significant and detailed clue to seeking new WSMs, has so far been little investigated by experiments.

A 3D van der Waals compound MoTe$_2$ experiences a lattice distortion, i.e., structural phase transition from high-temperature monoclinic $1T'$ to low-temperature orthorhombic $T_d$ phase at a critical temperature $T_c \sim 250$ K, which is accompanied with the breaking of *spatial inversion symmetry* (see the crystal structures of the $1T'$ and $T_d$ phases in Figure 1a).[35,36] In the high-temperature monoclinic $1T'$ phase, MoTe$_2$ was theoretically predicted to be a candidate for three-dimensional (3D) second-order topological insulators (SOTI) which host topologically protected gapless states on the one-dimensional hinges, but have insulating states on the two-dimensional surfaces and in the 3D bulk.[37,38] Below the structural phase transition temperature $T_c$, MoTe$_2$, which exhibits the low-temperature spatial-inversion-symmetry-broken orthorhombic $T_d$ structure, was theoretically and experimentally identified as a type-II WSM with broken Lorentz invariance, open bulk Fermi surfaces and Weyl points at the boundary between electron and hole pockets.[9-15] Here, the structural phase transition in MoTe$_2$ serves as a bridge connecting the 3D SOTI and type-II WSM states. Up to now, the structural phase transition in MoTe$_2$ was effectively tuned by applying ultrashort laser pulses, reducing the film thickness, exerting mechanical pressure, or doping carriers,[39-43] which provides excellent paradigms for manipulating its electronic states. Previously theoretical study indicates that the net charge transfer from the intralayer bonding state around the Y point of the Brillouin zone to the interlayer antibonding states along the Γ-A direction *near Fermi energy* lowers the total energy of the bulk MoTe$_2$, which suggests that the electronic-kinetic-energy reduction plays a significant role in the occurrence of the monoclinic-to-orthorhombic phase transition and thus is one of the key factors for tuning the structural phase transition and the electronic states in this 3D van der Waals material.[44] However, the electronic-kinetic-energy reduction along the in-plane axes (i.e., *a*-axis and *b*-axis) and the out-of-plane direction (i.e., *c*-axis), which would provide detailed and important information for understanding and manipulating the structural phase transition in MoTe$_2$, has rarely been investigated by experiments. In addition, as one of the prerequisites for studying the contribution of the electronic-kinetic-energy reduction to the structural phase transition in MoTe$_2$, coherent electronic transport was previously revealed to exist within its *ab*-plane.[36] The scarcity of the measurement of the out-of-plane electronic transport in MoTe$_2$ not only hinders a complete understanding of the structural phase transition, but also acts as an obstacle for developing *vertical devices* used for infrared photodetection and imaging based on this 3D van der Waals material.[45-48] Therefore, it is highly desirable to study the nature of its out-of-plane electronic transport across the structural phase transition. Furthermore, considering that (i) the electronic kinetic energy is proportional to the Drude weight (i.e., spectral weight of the *low-energy* Drude component),[49-52] and (ii) the decrease in the electronic kinetic energy would cause the spectral-weight transfer from the *low-energy* region to the *high-energy* region due to the spectral-weight conservation law,[49] the anisotropy of the reduction of the electronic kinetic energy (or Drude weight) is expected to lead to the increase in the anisotropy of the inter-band optical absorption at high energies, which can result in the enhancement of linear dichroism—the difference between the optical absorptions corresponding to the linearly polarized electrical fields applied parallel and perpendicular to a crystalline axis.[53-55] Although linear dichroism is a key property of polarization-sensitive photonic devices which are significant for emerging sensing and communication applications,[53-55] the temperature evolution of linear dichroism in MoTe$_2$ were seldomly measured.

Optical spectroscopy is an efficient experimental technique for investigating the anisotropy of charge dynamics and linear dichroism in a material as it can probe the optical absorption of free charge carriers and inter-band excitations using linearly polarized incident light.[49-52, 56-65] Here, we used an *in situ* gold and aluminum overcoating technique to measure the *a*-axis and *b*-axis optical reflectance spectra (i.e., $R_a(\omega)$ and $R_b(\omega)$) of the MoTe$_2$ single crystals on a Fourier transform spectrometer with the electric field (*E*) of the incident light applied along the crystalline *a*-axis and *b*-axis. Since the area of the crystal surface parallel to the *c*-axis is quite small, we utilized a microscope system (Hyperion) combined with the Fourier transform spectrometer to obtain the *c*-axis optical reflectance spectra (i.e., $R_c(\omega)$) with *E* // *c*-axis (see the details about the linearly-polarized optical reflectance measurements in Experimental

Section). Our linearly-polarized optical reflectance measurements at different temperatures show that the *a*-axis, *b*-axis and *c*-axis optical conductivity spectra (i.e., $\sigma_{1a}(\omega)$, $\sigma_{1b}(\omega)$ and $\sigma_{1c}(\omega)$) exhibit the optical responses of free charge carriers—Drude components at low energies, which indicates that not only the in-plane but also the out-of-plane charge transports are coherent in MoTe$_2$. In addition, the coherent inter-layer and in-plane electronic transports revealed by optical spectroscopy are in agreement with the metallic electrical resistivities which were measured with the currents applied along the out-of-plane direction and the two in-plane axes (*a*-axis and *b*-axis), respectively. More importantly, the spectral weight (i.e., Drude weight) of the Drude components in the $\sigma_{1a}(\omega)$, rather than the Drude weights in the $\sigma_{1b}(\omega)$ and $\sigma_{1c}(\omega)$, decreases dramatically below $T_c$, which shows a highly anisotropic reduction of the Drude weight. The highly anisotropic reduction of the Drude weight here reveals that the loss of the electronic kinetic energy along *one* particular crystalline direction (i.e., *a*-axis) contributes significantly to the occurrence of the structural phase transition in this 3D WSM. Furthermore, below $T_c$, since (i) the spectral weight of the optical conductivity is transferred from the Drude component to the high-energy region and (ii) the ratio between the Drude weights in the $\sigma_{1a}(\omega)$ and $\sigma_{1b}(\omega)$ is reduced with decreasing temperature, the normalized linear dichroism (NLD) around 770 meV in $T_d$-MoTe$_2$ grows remarkably, and then reaches the value of ~ 0.68 at $T$ = 8 K, which is the *strongest* NLD among the reported type-II WSMs.

## 2. Results and Discussion

In Figure 1b, the resistivities (i.e., $\rho_a$, $\rho_b$ and $\rho_c$) of the MoTe$_2$ single crystals measured with the electrical currents applied parallel to the crystalline *a*-axis, *b*-axis and *c*-axis decrease as the temperature is lowered (see the details about the electrical resistivity measurements in the Experimental Section), which exhibits metallic in-plane and out-of-plane charge transports in this van der Waals materials. Furthermore, the large values of the ratio between the out-of-plane and in-plane resistivities, i.e., $\rho_c(8K)/\rho_a(8K) \approx 18.5$, $\rho_c(300K)/\rho_a(300K) \approx 16.5$, $\rho_c(8K)/\rho_b(8K) \approx 9.0$, and $\rho_c(300K)/\rho_b(300K) \approx 9.9$ (see Figure 1c and 1d), indicate a high charge-transport anisotropy between the out-of-plane and in-plane-axis directions in MoTe$_2$. Besides, after the structural phase transition, the ratio $\rho_c/\rho_b$ between the out-of-plane and *b*-axis resistivities decreases weakly, but the ratio $\rho_c/\rho_a$ between the out-of-plane and *a*-axis resistivities increases obviously, which implies a distinction between the evolutions of the *a*-axis and *b*-axis charge transports below $T_c$. Considering that (i) the resistivity of a conventional metal has the form $\rho = m^*/(n\tau e^2)$, where $m^*$ is the effective mass of charge carriers, $n$ is the carrier concentration, $\tau^{-1}$ is the scattering rate and $e$ is the elementary charge, and (ii) the Drude weight $S_{\text{Drude}}$ obtained by optical spectroscopy study is proportional to the ratio $n/m^*$ between the carrier concentration and the charge-carrier effective mass in the expression of the resistivity, the temperature evolution of the electrical anisotropy should be related to the temperature dependence of the Drude-weight anisotropy. Optical spectroscopy thus enables us to gain insights into the temperature evolution of the electrical anisotropy.

To study the temperature dependences of the charge transports along the *a*-axis, *b*-axis and *c*-axis, we first measured the $R_a(\omega)$, $R_b(\omega)$ and $R_c(\omega)$ of the MoTe$_2$ single crystals in the temperature range from 8 to 340 K. Figure 2a-c depicts the $R_a(\omega)$, $R_b(\omega)$ and $R_c(\omega)$ of the MoTe$_2$ single crystals at several typical temperatures in the energy range up to 2500 meV (see the reflectance spectra measured at $T$ = 320 K and 340 K in Figure S4 of the Supporting Information). Both above and below the structural phase transition temperature $T_c$ ~ 250 K, the $R_a(\omega)$, $R_b(\omega)$ and $R_c(\omega)$ at energies lower than 50 meV are close to unity and increase as the temperature is lowered, which shows the optical response of free charge carriers. To further identify the optical responses of free charge carriers, we obtained the real part (i.e., $\sigma_{1a}(\omega)$, $\sigma_{1b}(\omega)$ and $\sigma_{1c}(\omega)$) of the *a*-axis, *b*-axis and *c*-axis optical conductivity spectra via the Kramers–Kronig transformation of the measured $R_a(\omega)$, $R_b(\omega)$ and $R_c(\omega)$ (see the details about the Kramers-Kronig transformation in the Experimental Section). In Figure 2d-f, the $\sigma_{1a}(\omega)$, $\sigma_{1b}(\omega)$ and $\sigma_{1c}(\omega)$ of the MoTe$_2$ single crystals at different temperatures exhibit the upturn-like features—Drude components at energies lower than 50 meV, which indicates that not only the in-plane but also the out-of-plane transports of free charge carriers in MoTe$_2$ are coherent above and below $T_c$ ~ 250 K. Herein, the coherence of the out-of-plane transports of free charge carriers in MoTe$_2$ revealed by the optical study is consistent with the metallic out-of-plane charge transports shown by the temperature dependence of the out-of-plane resistivity in Figure 1b. Moreover, compared with the $\sigma_{1b}(\omega)$ and $\sigma_{1c}(\omega)$, the $\sigma_{1a}(\omega)$ at low energies is significantly suppressed after cooling through the structural phase transition, which suggests that the spectral weight (i.e., Drude weight) of the Drude components in the $\sigma_{1a}(\omega)$ is very likely to be lost upon entering the type-II WSM state of $T_d$-MoTe$_2$.

To quantitatively investigate the temperature evolution of the Drude weights in the $\sigma_{1a}(\omega)$, $\sigma_{1b}(\omega)$ and $\sigma_{1c}(\omega)$ of MoTe$_2$, we fit the $\sigma_{1a}(\omega)$, $\sigma_{1b}(\omega)$ and $\sigma_{1c}(\omega)$ at different temperatures based on the Drude-Lorentz model (see the parameters of the Drude components and the Lorentzian peak around 770 meV for the Drude-Lorentz fit to the $\sigma_{1a}(\omega)$, $\sigma_{1b}(\omega)$ and $\sigma_{1c}(\omega)$ at different temperatures in Table 1 and the parameters of the other Lorentzian peaks for the Drude-Lorentz fit in Table S1-8 of the Supporting Information). Figure 3a-f show that the low-energy parts of the $\sigma_{1a}(\omega)$, $\sigma_{1b}(\omega)$ and $\sigma_{1c}(\omega)$ at $T$ = 8 K and 300 K can be well reproduced by two Drude components, respectively (see the Drude components for the Drude-Lorentz fit to the $\sigma_{1a}(\omega)$ and $\sigma_{1b}(\omega)$ at $T$ = 75 K, 150 K, 210 K, 270 K, 320 K and 340 K in Figure S5 of the Supporting Information and the Drude components for the Drude-Lorentz fit to the $\sigma_{1c}(\omega)$ at $T$ = 75 K, 150 K, 210 K and 270 K in Figure S6 of the Supporting Information). The obtained Drude weights $S_{Drude}$ in the $\sigma_{1a}(\omega)$, $\sigma_{1b}(\omega)$ and $\sigma_{1c}(\omega)$ were plotted as a function of temperature in Figure 3g-i. It is worth noticing that after the structural phase transition, the $S_{Drude}$ in the $\sigma_{1b}(\omega)$ is weakly decreased from ~ 2.5 × 10$^6$ cm$^{-1}$ to ~ 2.2 × 10$^6$ cm$^{-1}$ and the $S_{Drude}$ in the $\sigma_{1c}(\omega)$ are slightly reduced from ~ 1.4 × 10$^6$ cm$^{-1}$ to ~ 1.2 × 10$^6$ cm$^{-1}$, while the $S_{Drude}$ in the $\sigma_{1a}(\omega)$ is significantly suppressed from ~ 5.0 × 10$^6$ cm$^{-1}$ to ~ 3.3 × 10$^6$ cm$^{-1}$. The suppression (~ 1.7 × 10$^6$ cm$^{-1}$) of the $S_{Drude}$ in the $\sigma_{1a}(\omega)$ is about 6-8 times larger than the reduction of the $S_{Drude}$ in the $\sigma_{1b}(\omega)$ and the $S_{Drude}$ in the $\sigma_{1c}(\omega)$. Since (i) the resistivity in a conventional metallic state can be expressed as $\rho = m^*/(e^2 n\tau)$ (here $m^*$ is the effective mass of charge carriers, $e$ is the elementary charge, $n$ is the carrier concentration and $\tau^{-1}$ is the scattering rate), and (ii) the Drude weight $S_{Drude} \propto n/m^*$, the significant suppression of the $S_{Drude}$ in the $\sigma_{1a}(\omega)$, the weak decrease in the $S_{Drude}$ in the $\sigma_{1b}(\omega)$ and the slight reduction of the $S_{Drude}$ in the $\sigma_{1c}(\omega)$ are very likely to cause the obvious enhancement of the ratio $\rho_c/\rho_a$ and the weak reduction of the ratio $\rho_c/\rho_b$ after the structural phase transition. More importantly, considering the definition of the electronic kinetic energy in the optical investigation of a multiband material LaOFeP, we can get the linear relationship between the electronic kinetic energy (i.e., $K$) and the Drude weight: $K = 2\hbar^2 d_0 S_{Drude}/\pi e^2$, where $\hbar$ is Planck's constant divided by $2\pi$, and $d_0$ is the inter-MoTe$_2$-layer distance.[50] Therefore, the dramatic suppression of the $S_{Drude}$ in the $\sigma_{1a}(\omega)$, rather than the $S_{Drude}$ in the $\sigma_{1b}(\omega)$ and the $S_{Drude}$ in the $\sigma_{1c}(\omega)$, reveals that the loss of the electronic kinetic energy along *one* crystalline direction (i.e., *a*-axis contributes *significantly* to the occurrence of the structural phase transition in this 3D WSM.

At $T$ = 300 K > $T_c$, the in-plane Drude weights $S_{Drude}$ in the $\sigma_{1a}(\omega)$ and the $S_{Drude}$ in the $\sigma_{1b}(\omega)$ exhibit a ratio of ~ 2.0. After cooling through the structural phase transition, the ratio between the $S_{Drude}$ in the $\sigma_{1a}(\omega)$ and the $S_{Drude}$ in the $\sigma_{1b}(\omega)$ decreases and then reaches the value of ~ 1.4 at $T$ = 8 K. According to the spectral-weight conservation law, the lost Drude weight at low energies would be transferred to the *high-energy* optical-conductivity peaks. Thus, across the structural phase transition, the reduction of the ratio between the $S_{Drude}$ in the $\sigma_{1a}(\omega)$ and the $S_{Drude}$ in the $\sigma_{1b}(\omega)$ is expected to result in the enhancement of the ratio between the high-energy spectral weights in the $\sigma_{1a}(\omega)$ and the $\sigma_{1b}(\omega)$. Because the $\sigma_1(\omega)$ is proportional to the optical absorption, the possible growth of the ratio between the high-energy spectral weights in the $\sigma_{1a}(\omega)$ and the $\sigma_{1b}(\omega)$ implies the increase in the anisotropy of the high-energy optical absorption.

Noticeably, in Figure 2a and 2b, the $R_a(\omega)$ shows two distinct peak-like features around 770 meV and 1500 meV, respectively, while the $R_b(\omega)$ displays two dip-like features around 770 meV and 1500 meV. Moreover, below $T_c$, the peak-like features around 770 meV and 1500 meV in the $R_a(\omega)$ are remarkably enhanced, but the corresponding dip-like features in the $R_b(\omega)$ are continuously suppressed. The peak-like features in the $R_a(\omega)$ and the dip-like features in the $R_b(\omega)$, which not only are present in the same energy ranges, respectively but also have opposite temperature dependences of the intensities, suggest the existence of a strong and temperature-tunable optical anisotropy in MoTe$_2$. To quantitatively study the anisotropy of the in-plane optical absorption at high energies, we employed the normalized linear dichroism (NLD).[53,54] The NLD describes the difference between the optical absorptions corresponding to the linearly polarized electrical fields applied parallel and perpendicular to a crystalline axis and has the form:

$$\mathbf{NLD} = \frac{k_a - k_b}{k_a + k_b} \quad (1)$$

where $\boldsymbol{k_a}$ and $\boldsymbol{k_b}$ are the extinction coefficient of the complex refractive index $N_a$ and $N_b$ (i.e., $N = n + ik$) along the *a*-axis and *b*-axis of the MoTe$_2$ single crystals, respectively.[53,54] Here, the NLD varies within the range from −1 to 1. According to the Kramers-Kronig transformations of the measured $R(\omega)$ and the relationship between the complex refractive index and the complex reflectance (see the related details in the Methods and the derived extinction coefficient $\boldsymbol{k_a}$ and $\boldsymbol{k_b}$ of the MoTe$_2$ single crystals at different temperatures in Figure S8 of the Supporting Information), we obtained the photon-energy dependence of the NLD at different temperatures,[53-56] as displayed in Figure 4a. The NLD spectrum at each temperature shows two peak-like features and has the maximal value at ~ 770 meV. As displayed in Figure 4b, the maximal value of the NLD spectrum at ~ 770 meV exhibits a weak dependence on temperature in the monoclinic 1*T'* structure at $T > 270$ K, but increases remarkably from ~ 0.48 to ~ 0.68 in the type-II WSM phase of the orthorhombic $T_d$ structure as the temperature is lowered from 270 to 8 K. Furthermore, we plotted the temperature dependence of the in-plane optical absorption $(\boldsymbol{k_b} - \boldsymbol{k_a})/(\boldsymbol{k_a} + \boldsymbol{k_b})$ of MoTe$_2$. The $(\boldsymbol{k_b} - \boldsymbol{k_a})/(\boldsymbol{k_a} + \boldsymbol{k_b})$ spectra in the inset of Figure 4a show the peak-like features around 1100 meV and 1800 meV, respectively. The intensities of the peak-like features around 1100 meV and 1800 meV in the $(\boldsymbol{k_b} - \boldsymbol{k_a})/(\boldsymbol{k_a} + $ $k_b$) spectra show weak dependences on temperature in the monoclinic 1T' structure at T > 270 K, but increase separately to ~ 0.20 and ~ 0.28 as the temperature decreases from 270 K to 8 K. However, the maximal values (~ 0.20 and ~ 0.28) of the peak-like features at ~ 1100 meV and ~ 1800 meV in the $(k_b - k_a)/(k_a + k_b)$ spectra are much smaller than the maximal value (~ 0.68) of the normalized linear dichroism $(k_a - k_b)/(k_a + k_b)$ spectra at ~ 770 meV.

To gain insight into the enhancement of the in-plane optical-absorption anisotropy around 770 meV after the monoclinic-to-orthorhombic phase transition in MoTe$_2$, we analyzed the temperature evolution of the spectral weights of the high-energy-optical-absorption-induced Lorentzian peaks because of the linear relationship between the $\sigma_1(\omega)$ and the optical absorption. Herein, the spectral weights of the Lorentzian peaks are obtained by the Drude-Lorentz fit to the $\sigma_{1a}(\omega)$ and the $\sigma_{1b}(\omega)$ at different temperatures. Figure 4c and 4d show that (i) no Lorentzian peak is present around 770 meV in the $\sigma_{1b}(\omega)$ at T = 8 K and 300 K, and (ii) the change of the $\sigma_{1b}$ ($\omega$ = 770 meV) is slight as the temperature decreases from 300 to 8 K. On the contrary, Figure 4e and 4f display that the blue shaded Lorentzian peaks can be observed around 770 meV in the $\sigma_{1a}(\omega)$ at T = 8 K and 300 K. Moreover, the spectral weight of the blue shaded Lorentzian peak at ~ 770 meV at T = 8 K in Figure 4e is distinctly larger than that of the blue shaded Lorentzian peak at T = 300 K in Figure 4f (see the Lorentzian components for the Drude-Lorentz fit to the $\sigma_{1a}(\omega)$ and $\sigma_{1b}(\omega)$ at T = 75 K, 150 K, 210 K, 270 K, 320 K and 340 K in Figure S7 of the Supporting Information). In addition, as shown in Figure 4b, the spectral weight of the blue shaded Lorentzian peak at ~ 770 meV is significantly enhanced after cooling from 270 to 8 K, which is consistent with the temperature evolution of the maximal value of the NLD spectrum at ~ 770 meV (It is worth noticing that due to the coexistence of the orthorhombic $T_d$ structure and the monoclinic 1T' structure in the temperature range from 230 K to 290 K,[13,35,36,41,66,67] the enhancement of the NLD or spectral weight of Lorentzian peak should be present at $T \lesssim 290$ K, which ensures that the NLD or spectral weight of Lorentzian peak is significantly enhanced after cooling from 270 K to 8 K. However, it is a pity that we measured the optical spectra of MoTe$_2$ at the two temperatures (270 K and 300 K) below and above 290 K, which makes the NLD or spectral weight of Lorentzian peak seem to increase from 300 K in Figure 4b). As mentioned above, the enhancement of the spectral weight of the blue shaded Lorentzian peak at ~ 770 meV in the $\sigma_{1a}(\omega)$ results from the remarkable loss of the Drude weight at low energies due to the spectral transfer. Thus, the significant spectral-weight transfer in the $\sigma_{1a}(\omega)$, together with the absence of the Lorentzian peak around 770 meV in the $\sigma_{1b}(\omega)$ and the slight change of the $\sigma_{1b}$ ($\omega$ = 770 meV), causes the increase in the in-plane optical absorption anisotropy around 770 meV after the structural phase transition in MoTe$_2$.

Linearly polarized optical spectroscopy employed here is an efficient experimental tool for studying the charge dynamics and inter-band transitions of anisotropic materials. Herein, the Kramers-Kronig transformation of the reflectivity measured by linearly polarized optical spectroscopy is needed to obtain the corresponding imaginary part, i.e., the phase angle. Then, the other optical constants, such as complex refractive index and complex optical conductivity, can be obtained via the relationships between the complex reflectivity and the optical constants. It is worth noticing that Kramers-Kronig transformation cannot be simply applied to anisotropic materials because the contributions from different crystalline directions will mix.[68,69] Our optical study is mainly about the enhanced linear-dichroism and highly anisotropic Drude-weight-reduction in the Weyl semimetal $T_d$-MoTe$_2$ with *orthorhombic* crystal structure (space group of P2$_1$/m).[35] Above the structural phase transition temperature, MoTe$_2$ exhibits the monoclinic 1$T'$ crystal structure with a small tilting angle (~ 4˚) from the stacking direction (*c*-axis). Thus, the linear-dichroism and Drude weight reported here, which were obtained by our linearly polarized reflectivity measurements and the Kramers-Kronig transformation, are expected to be quite close to their intrinsic values.

Noteworthily, Mueller matrix ellipsometry (MME) is another powerful technique for investigating optical and dielectric properties of anisotropic materials.[70,71] Moreover, the MME can yield inherent tensor values of optical constants along a specific axis and thus can be used for quantitative studies of optical and electronic anisotropies in low-dimensional materials. In order to check whether the Kramers-Kronig transformation of the reflectivity spectra of MoTe$_2$ is available, we performed the variable-angle spectroscopic ellipsometry measurements at 300K (see the details about the spectroscopic ellipsometry measurements in the Experimental Section). Figure 5a and 5b show the extinction coefficient $\boldsymbol{k_a}$ and $\boldsymbol{k_b}$ at 300K, respectively, which were gotten by the variable-angle spectroscopic ellipsometry measurements.[72] The gray curve in Figure 5c here displays the linear-dichroism $(\boldsymbol{k_a} - \boldsymbol{k_b})/(\boldsymbol{k_a} + \boldsymbol{k_b})$, which was derived from the extinction coefficient $k_a$ and $k_b$ in Figure 5a and 5b. The linear-dichroism (see the red curve in Figure 5c here), which was obtained via our linearly polarized reflectivity measurements and the Kramers-Kronig transformation, is very similar to the linear-dichroism gotten by the variable-angle spectroscopic ellipsometry measurements. Therefore, the enhanced linear-dichroism and highly anisotropic Drude-weight-reduction in MoTe$_2$, which is reported in our manuscript, should be quite close to their intrinsic values. The peak positions of NLD spectrum obtained by the ellipsometry are slightly different from that determined by K-K transformation method, which might due to the error from the fitting process in the spectroscopic ellipsometry data or the K-K transformation method used for the linearly polarized reflectance spectra. Here, the maximal value (~ 0.68) of the NLD of type-II WSM $T_d$-MoTe$_2$ at $T$ = 8 K is larger than those (~ 0.59, ~ 0.48 and ~ 0.04) of type-II WSMs WTe$_2$, WP$_2$ and LaAlGe at $T$ = 8 K and thus represents the strongest in-plane optical anisotropy among the reported type-II WSMs (see Figure 6 here, Section 1 and Figure S9 in Supporting Information).

## 3. Conclusion

In summary, we have performed the linearly-polarized optical spectroscopy and electrical transport studies of MoTe$_2$ at different temperatures. The $\sigma_{1a}(\omega)$, $\sigma_{1b}(\omega)$ and $\sigma_{1c}(\omega)$ exhibit the Drude components at low energies, which indicates that both the in-plane and the out-of-plane electronic transports are coherent in MoTe$_2$. Besides, the coherent out-of-plane and in-plane electronic transports revealed by optical spectroscopy are in agreement with the metallic electrical resistivities which were measured with the currents applied along the out-of-plane direction and the two in-plane axes (*a*-axis and *b*-axis), respectively. More importantly, the Drude weight in the $\sigma_{1a}(\omega)$, rather than the Drude weights in the $\sigma_{1b}(\omega)$ and $\sigma_{1c}(\omega)$, decreases dramatically below $T_c$, which indicates that the Drude-weight reduction is highly anisotropic in this WSM and suggests that the electronic-kinetic-energy loss along the crystalline *a*-axis contributes significantly to the occurrence of the structural phase transition. In addition, below $T_c$, since (i) the significant spectral weight transfer from the Drude weight to the high-energy region around 770 meV in the $\sigma_{1a}(\omega)$ and (ii) the slight change of the spectral weight around 770 meV in the $\sigma_{1b}(\omega)$, the normalized NLD around 770 meV in $T_d$-MoTe$_2$ grows remarkably, and then reaches the value of ~ 0.68 at $T$ = 8 K, which is the *strongest* NLD among the reported type-II WSMs. Our results not only provide a paradigm for understanding the occurrence of spatial inversion symmetry breaking in WSMs but also paves the way for the development of exotic photonic devices based on WSMs.

## 4. Experimental Section

**Growth and characterization of the single crystals**. The single crystals of MoTe$_2$, WTe$_2$, and LaAlGe studied in this paper were grown by flux method.[73-75] For the growth of the MoTe$_2$ single crystals, the molybdenum powders and tellurium lumps with a molar ratio of 1:20 were placed in an alumina crucible and then sealed in an evacuated quartz tube. The raw materials were heated to 1000 °C within 10 hours and kept at this temperature for 24 hours. Then, they were cooled slowly down to 900 °C within 100 hours. At 900 °C, the excess tellurium flux was removed via centrifugation to obtain the MoTe$_2$ single crystals.[73] For the growth of the WTe$_2$ single crystals, the tungsten powders and the excess tellurium powders with a molar ratio of 1:20 were placed in an alumina crucible and then sealed in an evacuated quartz tube. The raw materials were heated to 1000 °C within 10 hours and hold at this temperature for 5 hours. They were cooled down to 800 °C with a rate of 1 °C/h and then cooled down to 700 °C with a rate of 5 °C/h. At 700 °C, the excess tellurium flux was removed via centrifugation to obtain the WTe$_2$ crystals.[74] For the growth of the LaAlGe single crystals, the lanthanum powders, germanium lumps and aluminum shots with a molar ratio of 1:2:20 were placed in alumina crucible and then sealed in an evacuated quartz tube. The raw materials were heated to 1175 °C with a rate of 200 °C/h and kept for 2 hours. Then, they were cooled down to 700 °C with a rate of 30 °C/h. At 700 °C, the aluminum flux was spined off to obtain the shiny LaAlGe single crystals.[75] The WP$_2$ single crystals were prepared by chemical vapor transport method.[76] The tungsten trioxide, red phosphorus and the transport agent iodine were mixed and sealed in an evacuated quartz tube. The quartz tube was placed in a two-zone tube furnace. The transport reaction was kept with a temperature gradient of 970—850 °C for 12 days. After the transport reaction, the needle-like WP$_2$ single crystals were obtained at the cool end of the quartz tube. The crystal orientation of the grown single crystals was characterized by single-crystal X-ray diffractometer (BRUKER D8 VENTURE) with Mo-K$\alpha$ radiation (0.7 Å).

**Electrical resistivity measurements.** The electrical resistivity of the MoTe$_2$ single crystals were measured using a standard four-probe method in a Quantum Design physical property measurement system (PPMS) with the electric currents applied along the $a$-axis, $b$-axis and $c$-axis, respectively. In order to reduce the influence of the voltages and currents along the other directions as much as possible,[77] (i) for the four-point probe resistivity measurement with the electric currents applied along the $a$-axis, we measured the thin flake (thickness ~ 50 μm) of the MoTe$_2$ single crystal with a large ratio (~ 11) between the $a$-axis and $b$-axis lengths and aligned the four-probe electrodes (i.e., $I_+$, $V_+$, $V_-$ and $I_-$) along the $a$-axis, as shown in Figure S2a of Supporting Information, (ii) for the four-point probe resistivity measurement with the electric currents applied along the $b$-axis, we measured the thin flake (thickness ~ 50 μm) with a large ratio (~ 8) between the $b$-axis and $a$-axis lengths and aligned the four-probe electrodes along the $b$-axis, as displayed in Figure S2b of Supporting Information, and (iii) for the four-point probe resistivity measurement with the electric currents applied along the $c$-axis, we utilized a Corbino-shape-like electrode configuration to measure the interlayer resistivity, as shown in Figure S2c of Supporting Information. In Corbino-shape-like electrode configuration, except a small circle at the center of top (or bottom) face (i.e., *ab*-plane) of the sample, the rest of the sample face, which is covered by the silver paint, serves as an electric-current contact, while a contact for the voltage-drop measurement is inside the small circle on the top and bottom face of the MoTe$_2$ single crystal, respectively. The Corbino-shape-like electrode configuration has been used intensively for obtaining the interlayer resistivity.[78-80] Moreover, the *a*-axis and *b*-axis resistivities here, together with the ratio between the *a*-axis and *b*-axis resistivities, are consistent with the results reported by the previous studies.[36,81]

**Optical reflectance measurements.** The measurements of the linearly polarized reflectance spectra with the electric field (*E*) of the incident light applied along the crystalline *a*-axis and *b*-axis in the energy range up to 25000 cm$^{-1}$ were carried out using an *in situ* gold and aluminum overcoating technique on a Bruker Vertex 80v spectrometer. The reflectance spectra measured with *E* // *c*-axis were performed using a microscope system (Hyperion) combined with the Fourier transform spectrometer. A gold (or aluminum) thin film was evaporated onto the half of the crystal face parallel to the *c*-axis. Then, we measured the ratio between the light intensities from the crystal face parallel to the *c*-axis and the gold (or aluminum) film on the crystal face. Finally, the *c*-axis reflectance spectra were obtained by multiplying the ratio between the light intensities from the crystal face parallel to the *c*-axis and the gold (or aluminum) film on the crystal face with the reflectivity spectrum of gold (or aluminum). Therein, the linearly polarized light with the electric fields directly along a crystallographic axis of the single crystal was produced when the light passes through a linear polarizer (see the schematic in Figure S3 of the Supporting Information).

**Spectroscopic ellipsometry measurements.** The spectroscopic ellipsometry measurements were performed using a variable-angle spectroscopic ellipsometer (J. A. Woollam) to obtain the extinction coefficient $k_a$ and $k_b$ along the *a*-axis and *b*-axis of MoTe$_2$. The *ab*-plane of the MoTe$_2$ single crystal was measured with the light polarization states along the *a*-axis and *b*-axis, respectively. The energy range of the ellipsometric data is from 300 to 1000 nm. The schematic of the spectroscopic ellipsometry setup is shown in Figure S3b of the Supporting Information. The angle of incidence ($\theta_i$) is 55°. For the spectroscopic ellipsometry measurements, the azimuthal angle orientation ($\Phi_i$) on the *ab*-plane of MoTe$_2$ single crystal are 0° and 90°, which correspond to the polarization along *a*-axis ($\Phi_i = 0°$) and *b*-axis ($\Phi_i = 90°$) of the MoTe$_2$ single crystal, respectively. The spectroscopic ellipsometry data of Ψ and Δ can be given by the reflectance coefficient ratio of the complex Fresnel coefficients between different polarization states: [82,83]

$$r = \frac{r_p}{r_s} = tan(\Psi)e^{i\Delta} \tag{2}$$

where $r$ is the generic ratio between the complex Fresnel coefficients with the polarization parallel ($r_p$: p-polarized light) and perpendicular ($r_s$: s-polarized light) to the plane of the incident light, tan (Ψ) and Δ are the amplitude ratio of reflection and the phase difference of the p- and s-polarized light, respectively. The measured raw ellipsometry data of Ψ and Δ at incident angles ($\theta_i$) of 55° with polarization along a-axis ($\Phi_i = 0°$) and b-axis ($\Phi_i = 90°$) of the MoTe2 single crystal are depicted in Figure S13 of the Supporting Information. The extinction coefficient $k_a$ and $k_b$ were determined by fitting the raw ellipsometry data with the attached commercial WVASE 32 software.[83] The combined Drude-Lorentz models were selected to fit the obtained raw ellipsometry data and the fitted curves are shown in the Figure S13 of the Supporting Information.

**Kramers-Kronig transformation.** According to the Kramers-Kronig relations between the real and imaginary parts of optical response functions, the phase shift $\theta(\omega)$ of the reflected light relative to the incident light can be derived from the measured $R(\omega)$:[56]

$$\theta(\omega) = -\frac{\omega}{\pi} P \int_0^{+\infty} \frac{ln\ R(\omega') - ln\ R(\omega)}{\omega'^2 - \omega^2} d\omega' \tag{3}$$

where $P$ denotes the Cauchy principal value. According to the relationship between the complex refractive index and the complex reflectance, the real part $n(\omega)$ and the imaginary part $k(\omega)$ of the complex refractive index $\widehat{N}(\omega)$ can be given by:

$$n(\omega) = \frac{1 - R(\omega)}{1 + R(\omega) - 2\sqrt{R(\omega)}Cos(\theta(\omega))} \tag{4}$$

$$k(\omega) = \frac{-2\sqrt{R(\omega)}\sin(\theta(\omega))}{1 + R(\omega) - 2\sqrt{R(\omega)}Cos(\theta(\omega))} \tag{5}$$

Furthermore, the real part $\sigma_1(\omega)$ of the optical conductivity have the following relationships with $n(\omega)$ and $k(\omega)$:

$$\sigma_1(\omega) = \frac{\omega n(\omega)k(\omega)}{2\pi} \tag{6}$$

Thus, the real part $\sigma_1(\omega)$ of the optical conductivity, the real part $n(\omega)$ and the imaginary part $k(\omega)$ of the refractive index can be obtained from the measured reflectance spectra $R(\omega)$.

**Supporting Information**

Supporting Information is available from the Wiley Online Library or from the corresponding author.

**Acknowledgements**

Z.-G.C. conceived and supervised this project. B.S. and W.L. carried out the optical measurements. W.W., J.Z. and S.A.Y. did theoretical analysis. B.S. carried out the transport measurements. G.G. and Q.L. grew the $MoTe_2$ single crystals. B.S. grew the $WP_2$ single crystals. X.D. and Y.S. grew the LaAlGe and $WTe_2$ single crystals. Z.-G.C., B.S. and J.L. analyzed the data. Z.-G.C. and B.S. wrote the paper. The authors acknowledge support from the Guangdong Basic and Applied Basic Research Foundation (Projects No. 2021B1515130007), the strategic Priority Research Program of Chinese Academy of Sciences (Project No. XDB33000000), the National Key Research and Development Program of China (Grant No. 2022YFA1403800), the National Natural Science Foundation of China (Projects Nos. 12022412, U21A20432, U2032204 and U22A6005), and the Synergetic Extreme Condition User Facility (SECUF)-Infrared Unit in THz and Infrared Experimental Station. The work at BNL was supported by the US Department of Energy, office of Basic Energy Sciences, contract no. DOE-sc0012704.

**Conflict of Interest**

The authors declare no conflict of interest.

References

[1]  X. Wan, A. M. Turner, A. Vishwanath, S. Y. Savrasov, *Phys. Rev. B* **2011**, *83*, 205101.

[2]  A. A. Burkov, L. Balents, *Phys. Rev. Lett.* **2011**, *107*, 127205.

[3]  H. Weng, C. Fang, Z. Fang, B. A. Bernevig, X. Dai, *Phys. Rev. X* **2015**, *5*, 011029.

[4]  S.-Y. Xu, I. Belopolski, N. Alidoust, M. Neupane, G. Bian, C. Zhang, R. Sankar, G. Chang, Z. Yuan, C.-C. Lee, S.-M. Huang, H. Zheng, J. Ma, D. S. Sanchez, B. Wang, A. Bansil, F. Chou, P. P. Shibayev, H. Lin, S. Jia, M. Z. Hasan, *Science* **2015**, *349*, 613.

[5] C. Shekhar, A. K. Nayak, Y. Sun, M. Schmidt, M. Nicklas, I. Leermakers, U. Zeitler, Y. Skourski, J. Wosnitza, Z. Liu, Y. Chen, W. Schnelle, H. Borrmann, Y. Grin, C. Felser, B. Yan, *Nat. Phys.* **2015**, *11*, 645.

[6] B. Q. Lv, H. M. Weng, B. B. Fu, X. P. Wang, H. Miao, J. Ma, P. Richard, X. C. Huang, L. X. Zhao, G. F. Chen, Z. Fang, X. Dai, T. Qian, H. Ding, *Phys. Rev. X* **2015**, *5*, 031013.

[7] L. X. Yang, Z. K. Liu, Y. Sun, H. Peng, H. F. Yang, T. Zhang, B. Zhou, Y. Zhang, Y. F. Guo, M. Rahn, D. Prabhakaran, Z. Hussain, S.-K. Mo, C. Felser, B. Yan, Y. L. Chen, *Nat. Phys*. **2015**, *11*, 728.

[8] A. A. Soluyanov, D. Gresch, Z. Wang, Q. Wu, M. Troyer, X. Dai, B. A. Bernevig, *Nat.* **2015**, *527*, 495.

[9] Y. Sun, S.-C. Wu, M. N. Ali, C. Felser, B. Yan, *Phys. Rev. B* **2015**, *92*, 161107(R).

[10] Z. Wang, D. Gresch, A. A. Soluyanov, W. Xie, S. Kushwaha, X. Dai, M. Troyer, R. J. Cava, B. A. Bernevig, *Phys. Rev. Lett.* **2016**, *117*, 056805.

[11] L. Huang, T. M. McCormick, M. Ochi, Z. Zhao, M.-T. Suzuki, R. Arita, Y. Wu, D. Mou, H. Cao, J. Yan, N. Trivedi, A. Kaminski, *Nat. Mater*. **2016**, *15*, 1155.

[12] K. Deng, G. Wan, P. Deng, K. Zhang, S. Ding, E. Wang, M. Yan, H. Huang, H. Zhang, Z. Xu, J. Denlinger, A. Fedorov, H. Yang, W. Duan, H. Yao, Y. Wu, S. Fan, H. Zhang, X. Chen, S. Zhou, *Nat. Phys*. **2016**, *12*, 1105.

[13] K. Zhang, C. Bao, Q. Gu, X. Ren, H. Zhang, K. Deng, Y. Wu, Y. Li, J. Feng, S. Zhou, *Nat. Commun.* **2016**, *7*, 13552.

[14] A. Tamai, Q. S. Wu, I. Cucchi, F. Y. Bruno, S. Riccò, T. K. Kim, M. Hoesch, C. Barreteau, E. Giannini, C. Besnard, A. A. Soluyanov, F. Baumberger, *Phy. Rev. X* **2016**, *6*, 031021.

[15] J. Jiang, Z. K. Liu, Y. Sun, H. F. Yang, C. R. Rajamathi, Y. P. Qi, L. X. Yang, C. Chen, H. Peng, C-C. Hwang, S. Z. Sun, S-K. Mo, I. Vobornik, J. Fujii, S. S. P. Parkin, C. Felser, B. H. Yan, Y. L. Chen, *Nat. Commun.* **2017**, *8*, 13973.

[16] G. Autès, D. Gresch, M. Troyer, A. A. Soluyanov, O. V. Yazyev, *Phys. Rev. Lett.* **2016**, *117*, 066402.

[17] S.-Y. Xu, N. Alidoust, G. Chang, H. Lu, B. Singh, I. Belopolski, D. S. Sanchez, X. Zhang, G. Bian, H. Zheng, M.-A. Husanu, Y. Bian, S.-M. Huang, C.-H. Hsu, T.-R. Chang, H.-T. Jeng, A. Bansil, T. Neupert, V. N. Strocov, H. Lin, S. Jia, M. Z. Hasan, *Sci. Adv.* **2017**, *3*, e1603266.

[18] P. Li, Y. Wen, X. He, Q. Zhang, C. Xia, Z.-M. Yu, S. A. Yang, Z. Zhu, H. N. Alshareef, X.-X. Zhang, *Nat. Commun.* **2017**, *8*, 2150.

[19] M.-Y. Yao, N. Xu, Q. S. Wu, G. Autès, N. Kumar, V. N. Strocov, N. C. Plumb, M. Radovic, O. V. Yazyev, C. Felser, J. Mesot, M. Shi, *Phys. Rev. Lett.* **2019**, *122*, 176402.

[20] G. Xu, H. Weng, Z. Wang, X. Dai, Z. Fang, *Phys. Rev. Lett.* **2011**, *107*, 186806.

[21] X. Huang, L. Zhao, Y. Long, P. Wang, D. Chen, Z. Yang, H. Liang, M. Xue, H. Weng, Z. Fang, X. Dai, G. Chen, *Phys. Rev. X* **2015**, *5*, 031023.

[22] M. Hirschberger, S. Kushwaha, Z. Wang, Q. Gibson, S. Liang, C. A. Belvin, B. A. Bernevig, R. J. Cava, N. P. Ong, *Nat. Mater.* **2016**, *15*, 1161.

[23] Y. Sun, Y. Zhang, C. Felser, B. Yan, *Phys. Rev. Lett.* **2016**, *117*, 146403.

[24] E. Liu, Y. Sun, N. Kumar, L. Muechler, A. Sun, L. Jiao, S.-Y. Yang, D. Liu, A. Liang, Q. Xu, J. Kroder, V. Süß, H. Borrmann, C. Shekhar, Z. Wang, C. Xi, W. Wang, W. Schnelle, S. Wirth, Y. Chen, S. T. B. Goennenwein, C. Felser, *Nat. Phys.* **2018**, *14*, 1125.

[25] A. A. Zyuzin, A. A. Burkov, *Phys. Rev. B* **2012**, *86*, 115133.

[26] C.-X. Liu, P. Ye, X.-L. Qi, *Phys. Rev. B* **2013**, *87*, 235306.

[27] M. M. Vazifeh, M. Franz, *Phys. Rev. Lett.* **2013**, *111*, 027201.

[28] C.-L. Zhang, S.-Y. Xu, I. Belopolski, Z. Yuan, Z. Lin, B. Tong, G. Bian, N. Alidoust, C.-C. Lee, S.-M. Huang, T.-R. Chang, G. Chang, C.-H. Hsu, H.-T. Jeng, M. Neupane, D. S. Sanchez, H. Zheng, J. Wang, H. Lin, C. Zhang, H.-Z. Lu, S.-Q. Shen, T. Neupert, M. Z. Hasan, S. Jia, *Nat. Commun.* **2016**, *7*, 10735.

[29] J. Lai, X. Liu, J. Ma, Q. Wang, K. Zhang, X. Ren, Y. Liu, Q. Gu, X. Zhuo, W. Lu, Y. Wu, Y. Li, J. Feng, S. Zhou, J.-H. Chen, D. Sun, *Adv. Mater.* **2018**, *30*, 1707152.

[30] S. Chi, Z. Li, Y. Xie, Y. Zhao, Z. Wang, L. Li, H. Yu, G. Wang, H. Weng, H. Zhang, J. Wang, *Adv. Mater.* **2018**, *30*, 1801372.

[31] J. Ma, Q. Gu, Y. Liu, J. Lai, P. Yu, X. Zhuo, Z. Liu, J.-H. Chen, J. Feng, D. Sun, *Nat. Mater.* **2019**, *18*, 476.

[32] W. Zhou, J.Chen, H. Gao, T. Hu, S. Ruan, A. Stroppa, W. Ren, *Adv. Mater.* **2019**, *31*, 1804629.

[33] Q. Wang, J. Zheng, Y. He, J. Cao, X. Liu, M. Wang, J. Ma, J. Lai, H. Lu, S. Jia, D. Yan, Y. Shi, J. Duan, J. Han, W. Xiao, J.-H. Chen, K. Sun, Y. Yao, D. Sun. *Nat. Commun.* **2019**, *10*, 5736.

[34] X. Zhuo, J. Lai, P. Yu, Z. Yu, J. Ma, W. Lu, M. Liu, Z. Liu, D. Sun, *Light Sci. Appl*. **2021**, *10*, 101.

[35] R. Clarke, E. Marseglia, H. P. Hughes, *Philos. Mag.* B **1978**, *38*, 121.

[36] H. P. Hughes, R. H. Friend, *J. Phys. C: Solid State Phys.* **1978**, *11*, L103.

[37] F. Tang, H. C. Po, A. Vishwanath, X. G. Wan, *Nat. Phys.* **2019**, *15*, 470.

[38] Z. Wang, B. J. Wieder, J. Li, B. Yan, B. A. Bernevig, *Phys. Rev. Lett.* **2019**, *123*, 186401.

[39] M. Y. Zhang, Z. X. Wang, Y. N. Li, L. Y. Shi, D. Wu, T. Lin, S. J. Zhang, Y. Q. Liu, Q. M. Liu, J. Wang, T. Dong, N. L. Wang, *Phys. Rev. X*, **2019**, *9*, 021036.

[40] B. Su, Y. Huang, Y. Hou, J. Li, R. Yang, Y. Ma, Y. Yang, G. Zhang, X. Zhou, J. Luo, Z.-G. Chen, *Adv. Sci.* **2022**, *9*, 2101532.

[41] J. L. Hart, L. Bhatt, Y. Zhu, M.-G. Han, E. Bianco, S. Li, D. J. Hynek, J. A. Schneeloch, Y. Tao, D. Louca, P. Guo, Y. Zhu, F. Jornada, E. J. Reed, L. F. Kourkoutis, J. J. Cha, *Nat. Commun.* **2023**, *14*, 4803.

[42] C. Heikes, I-L. Liu, T. Metz, C. Eckberg, P. Neves, Y. Wu, L. Hung, P. Piccoli, H. Cao, J. Leao, J. Paglione, T. Yildirim, N. P. Butch, W. Ratcliff II, *Phys. Rev. Materials* **2018**, *2*, 074202.

[43] P. Li, J. Cui, J. Zhou, D. Guo, Z. Zhao, J. Yi, J. Fan, Z. Ji, X. Jing, F. Qu, C. Yang, L. Lu, J. Lin, Z. Liu, G. Liu, *Adv. Mater.* **2019**, *31*, 1904641.

[44] H.-J. Kim, S.-H. Kang, I. Hamada, Y.-W. Son, *Phys. Rev. B* **2017**, *95*, 180101.

[45] H, Xu, C, Guo, J. Zhang, W. Guo, C.-N. Kuo, C. S. Lue, W. Hu, L. Wang, G. Chen, A. Politano, X. Chen, W. Lu. *Small* **2019**, *15*, 1903362.

[46] M. S. Shawkat, T. A. Chowdhury, H.-S. Chung, S. Sattar, T.-J. Ko, J. A. Larssone, Y. Jung, *Nanoscale* **2020**, *12*, 23116.

[47] D. Wu, C. Guo, L. Zeng, X. Ren, Z. Shi, L. Wen, Q. Chen, M. Zhang, X. Li, C. Shan, J. Jie. *Light Sci Appl* **2023**, *12*, 5.

[48] X. Li, S. Wu, D. Wu, T. Zhao, P. Lin, Z. Shi, Y. Tian, X. Li, L. Zeng, X. Yu, *InfoMat*. **2024**, *6*, e12499.

[49] D. N. Basov, T. Timusk, *Rev. Mod. Phys.* **2005**, *77*, 721.

[50] M. M. Qazilbash, J. J. Hamlin, R. E. Baumbach, L. Zhang, D. J. Singh, M. B. Maple, D. N. Basov, *Nat. Phys.* **2009**, *5*, 647.

[51] Z. G. Chen, R. H. Yuan, T. Dong, N. L. Wang, *Phys. Rev. B* **2010**, *81*, 100502.

[52] Y. Xu, J. Zhao, C. Yi, Q. Wang, Q. Yin, Y. Wang, X. Hu, L. Wang, E. Liu, G. Xu, L. Lu, A. A. Soluyanov, H. Lei, Y. Shi, J. Luo, Z.-G. Chen. *Nat. Commun.* **2020**, *11*, 3985.

[53] S. Niu, G. Joe, H. Zhao, Y. Zhou, T. Orvis, H. Huyan, J. Salman, K. Mahalingam, B. Urwin, J. Wu, Y. Liu, T. E. Tiwald, S. B. Cronin, B. M. Howe, M. Mecklenburg, R. Haiges, D. J. Singh, H. Wang, M. A. Kats, J. Ravichandran, *Nat. Photon.* **2018**, *12*, 392.

[54] J. Wu, X. Cong, S. Niu, F. Liu, H. Zhao, Z. Du, J. Ravichandran, P.-H. Tan, H. Wang, *Adv. Mater.* **2019**, *31*, 1902118.

[55] Z. Guo, H. Gu, M. Fang, B. Song, W. Wang, X. Chen, C. Zhang, H. Jiang, L. Wang, S. Li, *ACS Materials Lett*, **2021**, *3*, 525.

[56] M. Dressel, G. Grüner, *Electrodynamics of Solids: Optical Properties of Electrons in Matter*, Cambridge University Press, Cambridge **2002**.

[57] Z. G. Chen, T. Dong, R. H. Ruan, B. F. Hu, B. Cheng, W. Z. Hu, P. Zheng, Z. Fang, X. Dai, N. L. Wang, *Phys. Rev. Lett.* **2010**, *105*, 097003.

[58] S. J. Moon, C. C. Homes, A. Akrap, Z. J. Xu, J. S. Wen, Z. W. Lin, Q. Li, G. D. Gu, D. N. Basov, *Phys. Rev. Lett.* **2011**, *106*, 217001.

[59] M. Nakajima, T. Liang, S. Ishida, Y. Tomioka, K. Kihou, C. H. Lee, A. Iyo, H. Eisaki, T. Kakeshita, T. Ito, S. Uchida, *Proc. Natl. Acad. Sci. USA* **2011**, *108*, 12238.

[60] C. Mirri, A. Dusza, S. Bastelberger, M. Chinotti, L. Degiorgi, J.-H. Chu, H.-H. Kuo, I. R. Fisher, *Phys. Rev. Lett.* **2015**, *115*, 107001.

[61] C. C. Homes, M. N. Ali, R. J. Cava, *Phys. Rev. B* **2015**, *92*, 161109.

[62] A. J. Frenzel, C. C. Homes, Q. D. Gibson, Y. M. Shao, K. W. Post, A. Charnukha, R. J. Cava, D. N. Basov, *Phys. Rev. B* **2017**, *95*, 245140.

[63] B. Su, Y. Song, Y. Hou, X. Chen, J. Zhao, Y. Ma, Y. Yang, J. Guo, J. Luo, Z.-G. Chen, *Adv. Mater.* **2019**, *31*, 1903498.

[64] S.-I. Kimura, Y. Nakajima, Z. Mita, R. Jha, R. Higashinaka, T. D. Matsuda, Y. Aoki, *Phys. Rev. B* **2019**, *99*, 195203.

[65] D. Santos-Cottin, E. Martino, F. Le Mardelé, C. Witteveen, F. O. von Rohr, C. C. Homes, Z. Rukelj, A. Akrap, *Phys. Rev. Mater.* **2020**, *4*, 021201.

[66] T. Zandt, H. Dwelk, C. Janowitz, R. Manzke, *J Alloy. Compd.* **2007**, *442*, 216.

[67] Y. Qi, P. G. Naumov, M. N. Ali, C. R. Rajamathi, W. Schnelle, O. Barkalov, M. Hanfland, S.-C. Wu, C. Shekhar, Y. Sun, V. Süß, M. Schmidt, U. Schwarz, E. Pippel, P. Werner, R. Hillebrand, T. Förster, E. Kampert, S. Parkin, R. J. Cava, C. Felser, B. Yan, S. A. Medvedev, *Nat. Commun.* **2016**, *7*, 11038.

[68] M. Dressel, B. Gompf, D. Faltermeier, A. K. Tripathi, J. Pflaum, M. Schubert, *Opt. Express* **2008**, *16*, 19770.

[69] C. Marcos, *Representation Surfaces of Optical Properties of Crystals*, Crystallography: Introduction to the Study of Minerals, Springer Nature, Switzerland **2022**, 287.

[70] M. Fang, H. Gu, Z. Guo, J. Liu, L. Huang, S. Liu, *Appl. Surf. Sci.* **2022**, *605*, 154813.

[71] Z. Guo, H. Gu, M. Fang, L. Ye, S. Liu, *Nanoscale* **2022**, *14*, 12238.

[72] G. E. Jellison Jr, J. S. Baba, *J. Opt. Soc. Am. A* **2006**, *23*, 468.

[73] Z. Guguchia, F. von Rohr, Z. Shermadini, A. T. Lee, S. Banerjee, A. R. Wieteska, C. A. Marianetti, B. A. Frandsen, H. Luetkens, Z. Gong, S. C. Cheung, C. Baines, A. Shengelaya, G. Taniashvili, A. N. Pasupathy, E. Morenzoni, S. J. L. Billinge, A. Amato, R. J. Cava, R. Khasanov, Y. J. Uemura, *Nat. Commun.* **2017**, *8*, 1082.

[74] D. Kang, Y. Zhou, W. Yi, C. Yang, J. Guo, Y. Shi, S. Zhang, Z. Wang, C. Zhang, S. Jiang, A. Li, K. Yang, Q. Wu, G. Zhang, L. Sun, Z. Zhao, *Nat. Commun.* **2015**, *6*, 7804.

[75] S. Bobev, P. H. Tobash, V. Fritsch, J. D. Thompson, M. F. Hundley, J. L. Sarrao, Z. Fisk, *J. Solid State Chem*. **2005**, *178*, 2091.

[76] H. Mathis, R. Glaum, R. Gruehn, *Acta Chem. Scand*. **1991**, *45*, 781.

[77] I. Miccoli, F. Edler, H. Pfnür, C. Tegenkamp, *J. Phys.: Condens. Matter* **2015**, *27*, 223201.

[78] C. Boulesteix, Y. Marietti, T. Badeche, H. Tatarenko-Zapolsky, V. Grachev, O. Monnereau, H. Faqir, G. Vacquier, *J. Phys. Chem. Solids* **2000**, *61*, 585.

[79] Y. Xiang, Q. Li, Y. Li, W. Xie, H. Yang, Z. Wang, Y. Yao, H. H. Wen, *Nat. Commun.* **2021**, *12*, 6727.

[80] X. F. Wang, T. Wu, G. Wu, H. Chen, Y. L. Xie, J. J. Ying, Y. J. Yan, R. H. Liu, X. H. Chen, *Phys. Rev. Lett*. **2009**, *102*, 117005.

[81] Y.-Y. Lv, X. Li, B. Pang, L. Cao, D. Lin, B.-B. Zhang, S.-H. Yao, Y. B. Chen, J. Zhou, S.-T. Dong, S.-T. Zhang, M.-H. Lu, Y.-F. Chen, *J. Appl. Phys.* **2017**, *122*, 045102.

[82] E. Garcia-Caurel, A. De Martino, J-P Gaston, Y. Li, Appl. *Spectrosc.* **2013**, *67*, 1.

[83] G. E. Jellison Jr, J. S. Baba, *J. Opt. Soc. Am. A* **2006**, *23*, 468.

**Figures, Figure Legends and Table**

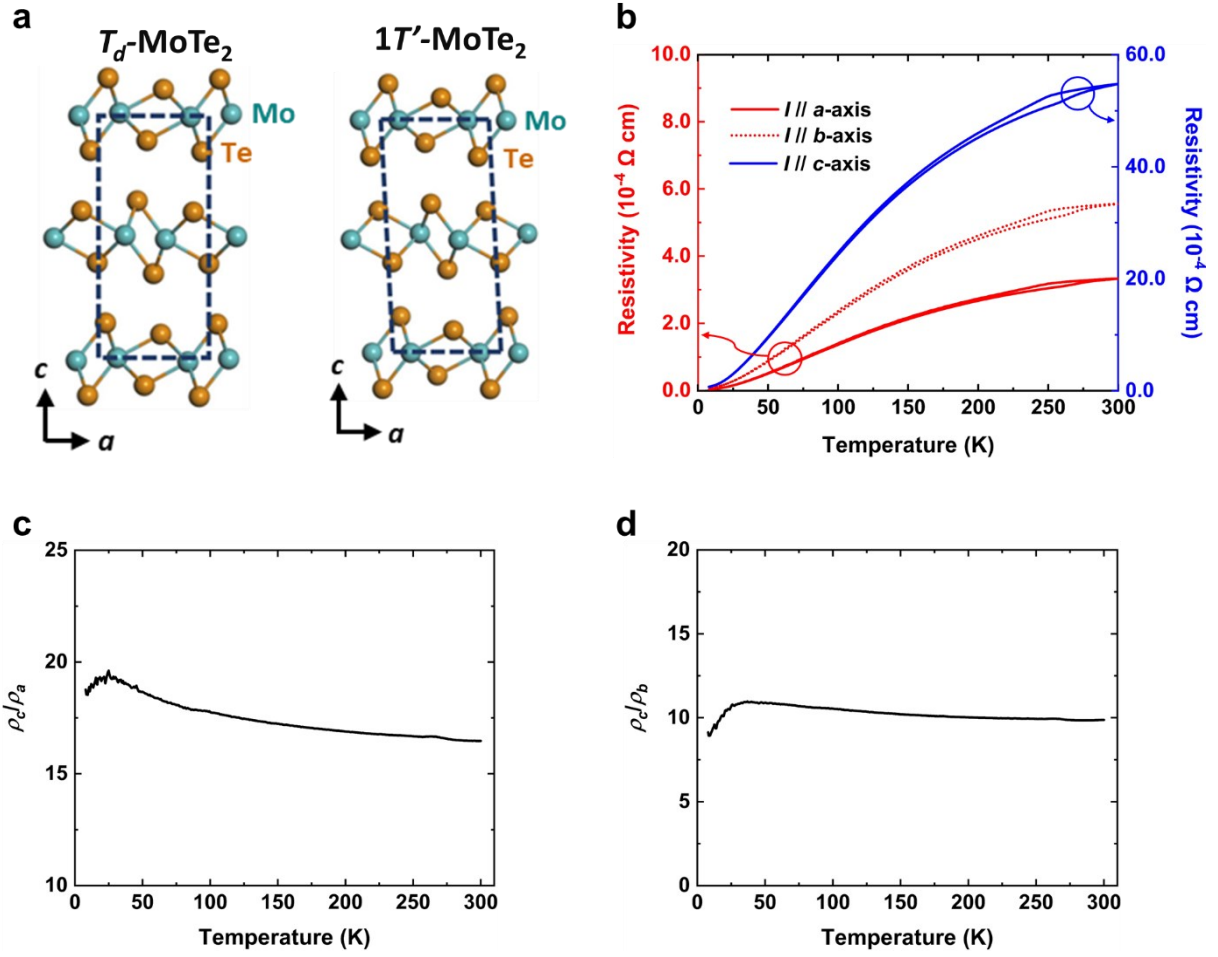

**Figure 1.** Crystal structures and resistivities of MoTe$_2$. a) Monoclinic 1$T'$ and orthorhombic $T_d$ structures of MoTe$_2$. The Mo and Te atoms are denoted by the green and brown spheres, respectively. b) Temperature dependence of the resistivities measured with the electric current (i.e., $I$) applied along the crystalline $a$-axis, $b$-axis and $c$-axis, respectively. Around the structural phase transition temperature ~ 250 K, the hysteretic features are present in the resistivity curves. c) Temperature dependent ratio $\rho_c/\rho_a$ between the $c$-axis resistivity and the $a$-axis resistivity. d) Temperature dependent ratio $\rho_c/\rho_b$ between the $c$-axis resistivity and the $b$-axis resistivity.

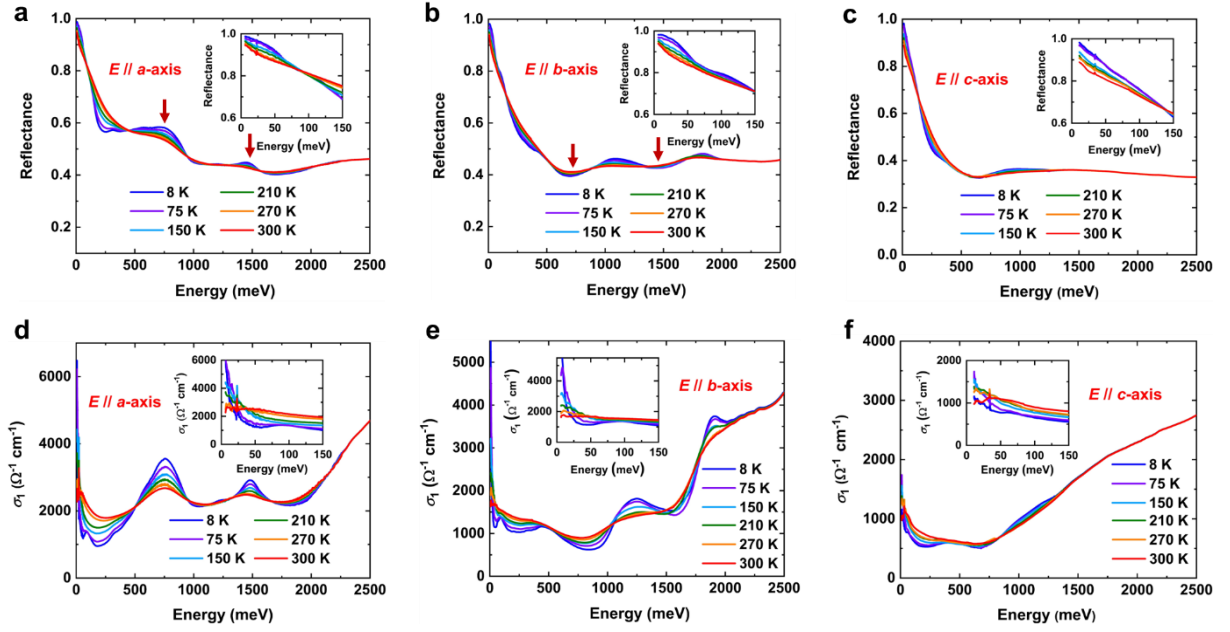

**Figure 2.** Optical response of MoTe$_2$ at different temperatures. a-c) Reflectance spectra of MoTe$_2$ measured with the electric field $E$ applied parallel to the $a$-axis, $b$-axis and $c$-axis, respectively. The insets in Figure a-c show the $a$-axis, $b$-axis and $c$-axis reflectance spectra below 150 meV. The two red arrows in a) indicate the two peak-like features around 770 meV and 1500 meV, respectively. The two red arrows in b) indicate the two dip-like features around 770 meV and 1500 meV, respectively. d-f) Real part (i.e., $\sigma_{1a}(\omega)$, $\sigma_{1b}(\omega)$ and $\sigma_{1c}(\omega)$) of the $a$-axis, $b$-axis and $c$-axis optical conductivity spectra. The insets in Figure d-f display the $a$-axis, $b$-axis and $c$-axis optical conductivity spectra below 150 meV.

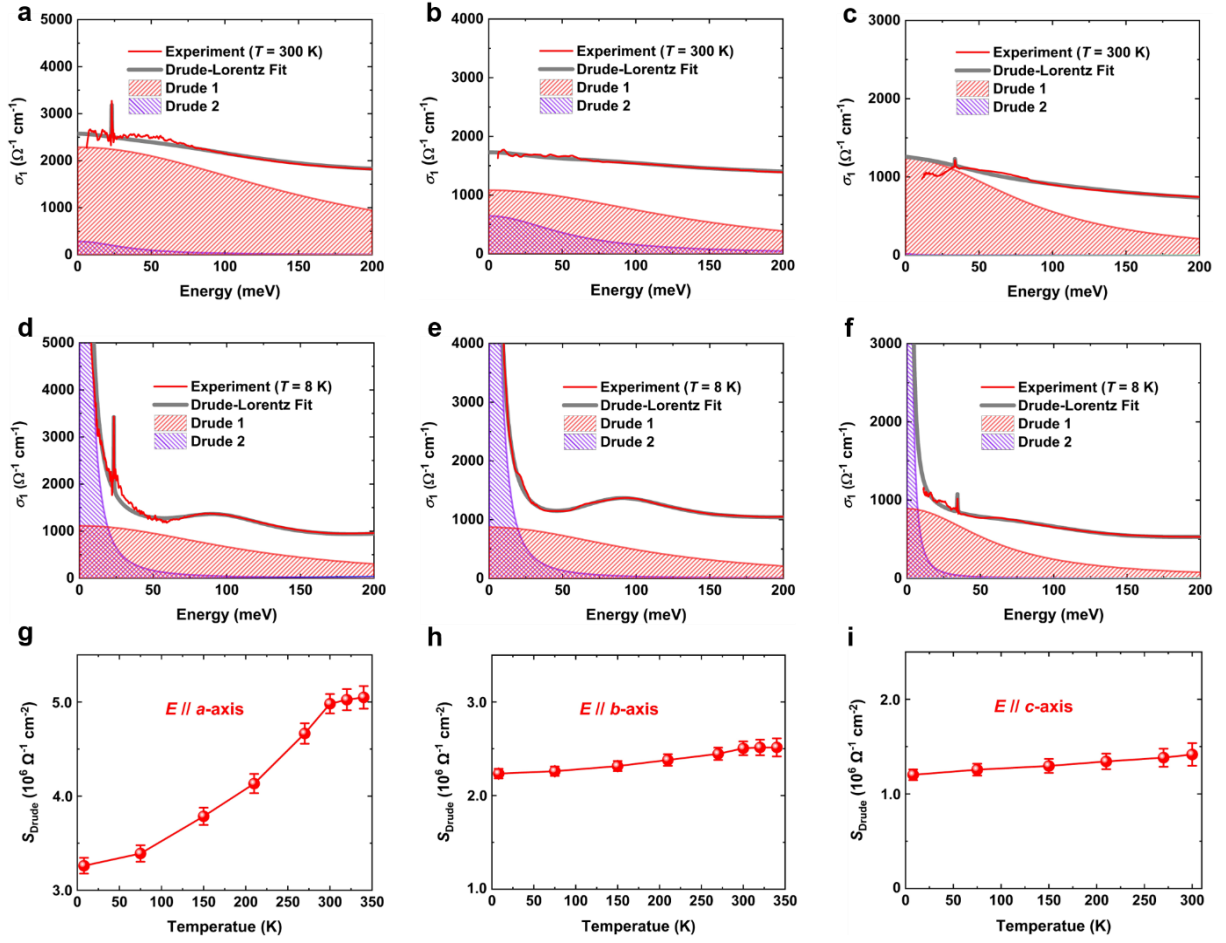

**Figure 3.** Drude weights in the $\sigma_{1a}(\omega)$, $\sigma_{1b}(\omega)$ and $\sigma_{1c}(\omega)$ of MoTe$_2$ at different temperatures. a-c) Drude components in the $\sigma_{1a}(\omega)$, $\sigma_{1b}(\omega)$ and $\sigma_{1c}(\omega)$ at $T$ = 300 K. d-f) Drude components in the $\sigma_{1a}(\omega)$, $\sigma_{1b}(\omega)$ and $\sigma_{1c}(\omega)$ at $T$ = 8 K. g-i) Temperature evolutions of the Drude weights in the $\sigma_{1a}(\omega)$, $\sigma_{1b}(\omega)$ and $\sigma_{1c}(\omega)$.

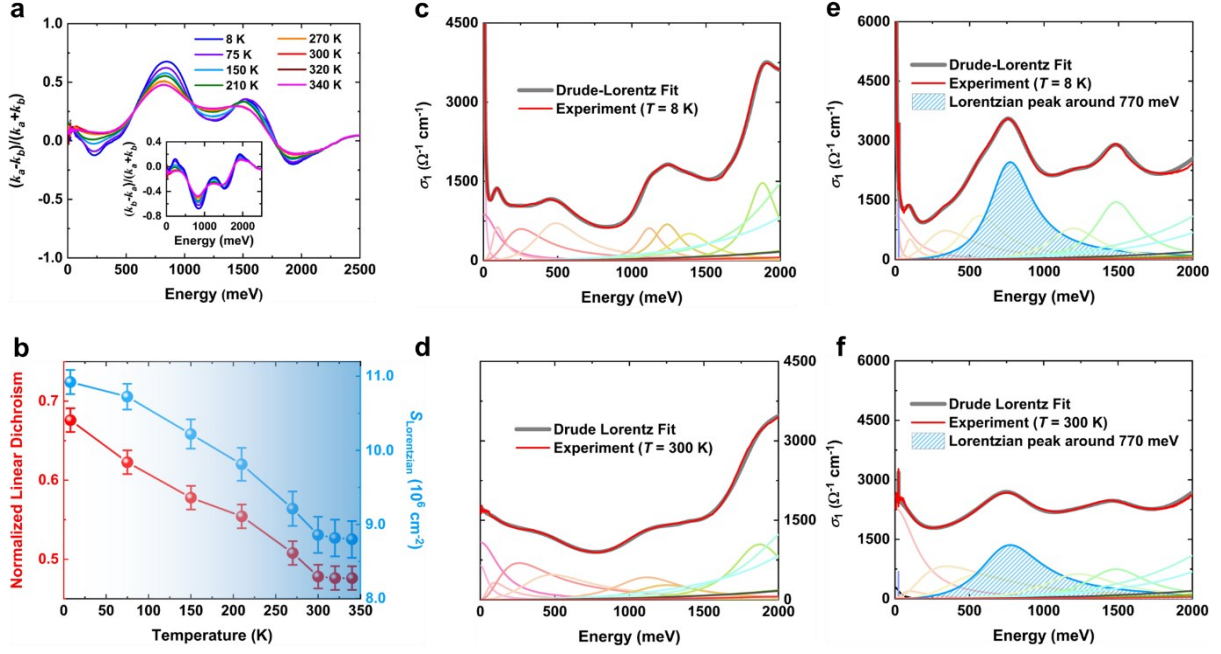

**Figure 4.** Normalized linear dichroism of MoTe$_2$ at different temperatures. a) Photon energy dependence of the normalized linear dichroism at different temperatures. The inset shows the temperature dependence of the in-plane optical absorption $(k_b - k_a)/(k_a + k_b)$ of MoTe$_2$ at different temperatures. b) Temperature dependence of the normalized linear dichroism at ~ 770 meV (see the red dots) and temperature evolution of the spectral weight of the Lorentzian peak around 770 meV for fitting the $\sigma_{1a}(\omega)$ (see the blue dots). c) and d) Drude components and Lorentzian peaks for fitting the $\sigma_{1b}(\omega)$ at $T$ = 8 K and 300 K. e) and f) Drude components and Lorentzian peaks for fitting the $\sigma_{1a}(\omega)$ at $T$ = 8 K and 300 K. The blue shaded areas in e) and f) indicate the Lorentzian peak around 770 meV for fitting the $\sigma_{1a}(\omega)$ at $T$ = 8 K and 300 K.

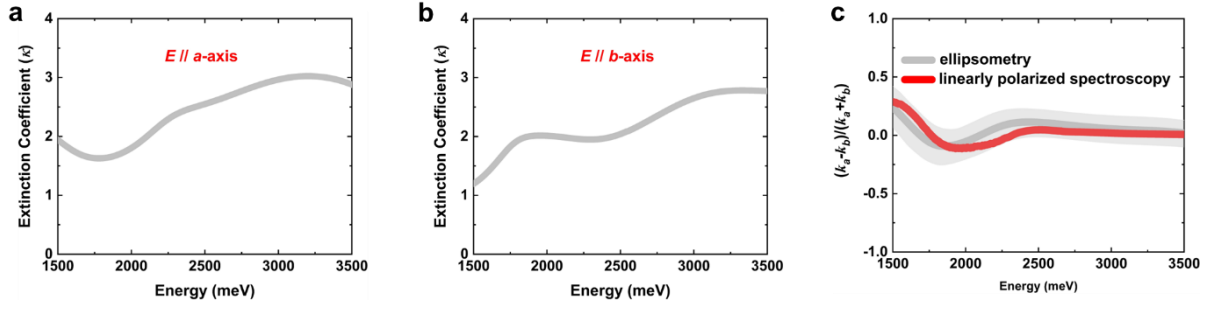

**Figure 5.** a**)** and b**)** Extinction coefficient $k_a$ and $k_b$ at 300K obtained by the variable-angle spectroscopic ellipsometry measurements along the *a*-axis and *b*-axis of the MoTe$_2$ single crystal, respectively. c) Normalized linear dichroism of MoTe$_2$ at 300 K gotten by the variable-angle spectroscopic ellipsometry measurements and linearly polarized spectroscopy measurements, respectively. The grey shadow indicates the error bar of the normalized linear dichroism of MoTe$_2$ at 300 K gotten by the variable-angle spectroscopic ellipsometry measurements.

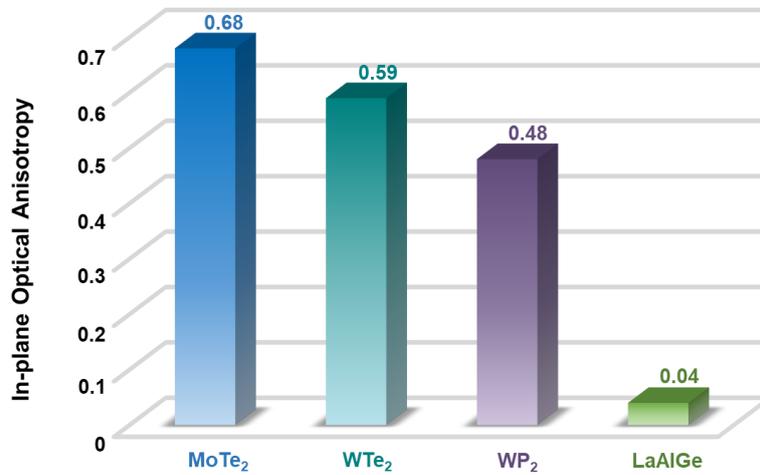

**Figure 6.** In-plane optical anisotropies (i.e., maximal values of the normalized linear dichroism spectra) of the reported type-II Weyl semimetals MoTe$_2$, WTe$_2$, WP$_2$ and LaAlGe.

**Table 1.** The parameters of the Drude components and the Lorentzian peak around 770 meV for the Drude-Lorentz fit to the $\sigma_{1a}(\omega)$, $\sigma_{1b}(\omega)$ and $\sigma_{1c}(\omega)$ at different temperatures.

| Temperature | Polarization | a-axis | | | b-axis | | | c-axis | | |
|---|---|---|---|---|---|---|---|---|---|---|
| | Parameters / Model | $\omega_0$ (meV) | $\omega_p$ (meV) | $1/\tau$ (meV) | $\omega_0$ (meV) | $\omega_p$ (meV) | $1/\tau$ (meV) | $\omega_0$ (meV) | $\omega_p$ (meV) | $1/\tau$ (meV) |
| 8 K | Drude term 1 | 0.00 | 951.42 | 3.41 | 0.00 | 761.88 | 4.36 | 0.00 | 547.53 | 1.19 |
| | Drude term 2 | 0.00 | 1005.89 | 122.05 | 0.00 | 856.33 | 112.95 | 0.00 | 637.89 | 61.46 |
| | Lorentzian term | 772.54 | 2533.94 | 351.30 | -- | -- | -- | -- | -- | -- |
| 75 K | Drude term 1 | 0.00 | 836.78 | 7.96 | 0.00 | 708.65 | 11.29 | 0.00 | 503.99 | 2.44 |
| | Drude term 2 | 0.00 | 1137.16 | 149.71 | 0.00 | 909.18 | 127.04 | 0.00 | 696.13 | 62.12 |
| | Lorentzian term | 774.87 | 2511.02 | 387.82 | -- | -- | -- | -- | -- | -- |
| 150 K | Drude term 1 | 0.00 | 791.15 | 15.12 | 0.00 | 647.50 | 19.89 | 0.00 | 272.47 | 5.41 |
| | Drude term 2 | 0.00 | 1258.35 | 163.15 | 0.00 | 970.47 | 136.42 | 0.00 | 829.10 | 75.30 |
| | Lorentzian term | 774.87 | 2451.19 | 425.57 | -- | -- | -- | -- | -- | -- |
| 210 K | Drude term 1 | 0.00 | 700.78 | 20.53 | 0.00 | 638.05 | 33.94 | 0.00 | 140.60 | 5.44 |
| | Drude term 2 | 0.00 | 1392.46 | 167.33 | 0.00 | 995.98 | 144.19 | 0.00 | 877.43 | 78.02 |
| | Lorentzian term | 774.87 | 2401.78 | 460.09 | -- | -- | -- | -- | -- | -- |
| 270 K | Drude term 1 | 0.00 | 496.11 | 25.66 | 0.00 | 569.82 | 42.58 | 0.00 | 60.66 | 5.44 |
| | Drude term 2 | 0.00 | 1580.03 | 167.52 | 0.00 | 1055.19 | 146.57 | 0.00 | 899.78 | 82.02 |
| | Lorentzian term | 774.87 | 2327.19 | 496.06 | -- | -- | -- | -- | -- | -- |
| 300 K | Drude term 1 | 0.00 | 276.64 | 36.02 | 0.00 | 519.50 | 56.23 | 0.00 | 31.39 | 6.38 |
| | Drude term 2 | 0.00 | 1688.63 | 167.64 | 0.00 | 1096.67 | 149.24 | 0.00 | 912.13 | 90.49 |
| | Lorentzian term | 774.87 | 2282.16 | 516.87 | -- | -- | -- | -- | -- | -- |
| 320 K | Drude term 1 | 0.00 | 228.60 | 37.83 | 0.00 | 472.40 | 61.31 | -- | -- | -- |
| | Drude term 2 | 0.00 | 1703.40 | 168.81 | 0.00 | 1120.03 | 149.86 | -- | -- | -- |
| | Lorentzian term | 774.86 | 2276.75 | 519.67 | -- | -- | -- | -- | -- | -- |
| 340 K | Drude term 1 | 0.00 | 210.20 | 39.07 | 0.00 | 455.87 | 64.41 | -- | -- | -- |
| | Drude term 2 | 0.00 | 1710.02 | 169.98 | 0.00 | 1127.20 | 150.28 | -- | -- | -- |
| | Lorentzian term | 774.86 | 2274.41 | 519.36 | -- | -- | -- | -- | -- | -- |